# Coherent ultra-broadband laser-assisted injection radiation from a laser plasma accelerator

B. Miao, L. Feder, J. Elle, A. J. Goers, D. Woodbury, F. Salehi, J. K. Wahlstrand, and H. M. Milchberg*

*Institute for Research in Electronics and Applied Physics, University of Maryland, College Park, MD 20742*

milch@umd.edu

**Abstract**: The injection of electrons into a laser wakefield accelerator (LWFA) is observed to generate an intense coherent ultra-broadband and ultrashort pulse radiation flash, consistent with the acceleration of electrons from rest to nearly the speed of light in a distance $< \sim 1\ \mu m$. The flash is sufficiently bright to induce large nonlinear refractive index shifts in optical materials; we estimate a source brightness temperature of $\sim 10^{18}$ K. We present measurements of the flash spectra, coherence, pulse duration, polarization and angular distribution, providing a detailed picture of electron injection dynamics in LWFA. These are characteristic of laser-assisted injection of off-axis electrons, which preserves wake coherence.

## 1. Introduction

The accelerating structure in a laser wakefield electron accelerator is provided by the relativistic nonlinear plasma wave driven by the ponderomotive force of the laser pulse. The plasma wave, which follows the pulse as a wake, can support axial electric fields >100 GV/m [1]. Electrons can be accelerated after self-injection of background plasma electrons or external injection of relativistic electrons into the wake structure. In the self-injection process [2–4], trapping of background electrons in the wake potential can lead to acceleration of high charge bunches from rest to nearly the speed of light over very short distances, generating radiation. At low plasma densities, this radiation has been observed to be weak and incoherent, even with a high-energy laser driver [5]. However, with near-critical plasma densities, using low energy millijoule-scale lasers, our preliminary experiments showed that injection generates an intense *coherent* broadband ($\Delta\lambda/\lambda \sim 1$) flash of radiation [6] owing to acceleration over distances short compared to the radiated wavelength. Flashes occur axially in the plasma at electron injection locations, which closely follow relativistic self-focusing beam collapse and nonlinear plasma wave generation. A flash is estimated to contain up to a few percent of the laser energy, and can overwhelm optical diagnostics such as probe laser beams [6]. In astrophysics, coherent radio wave emission from pulsars is an analogous process, as it may originate from extreme acceleration of dense electron bunches over distances much shorter than the radiated wavelength [7].

In this paper, we present detailed measurements showing that under our conditions of near critical density plasma the flash is an ultrashort pulse optical radiation source of at most few-cycle duration, whose coherence, pulsewidth, angular distribution, and polarization are consistent with synchrotron radiation emission from *laser-assisted* injection of dense electron bunches into the plasma wave accelerating structure. The emission is remarkably bright for a non-laser source, with a brightness temperature estimated to be $\sim 10^{18}$ K. At lower laser power, we observe the flash spectrum peaked at harmonics of the plasma frequency (the first measurement of such harmonics), a signature of sequential trapping in successive plasma wave potential buckets. At higher laser power, the flash spectrum is more continuous, characteristic of electron trapping largely in the main bucket following the laser pulse, with the following buckets strongly dissipated by beam loading.

Our analysis of injection flash radiation provides a detailed picture of the dynamics inside the laser plasma accelerator. If one considers "wave-breaking" to imply destruction of plasma wave coherence [8], our results lead us to identify the flash radiation as originating from laser-assisted injection rather than wave-breaking, as the radiation shows a characteristic polarization signature and coherent periodic features necessitating plasma wave survival through the injection process.

## 2. Flash experimental setup

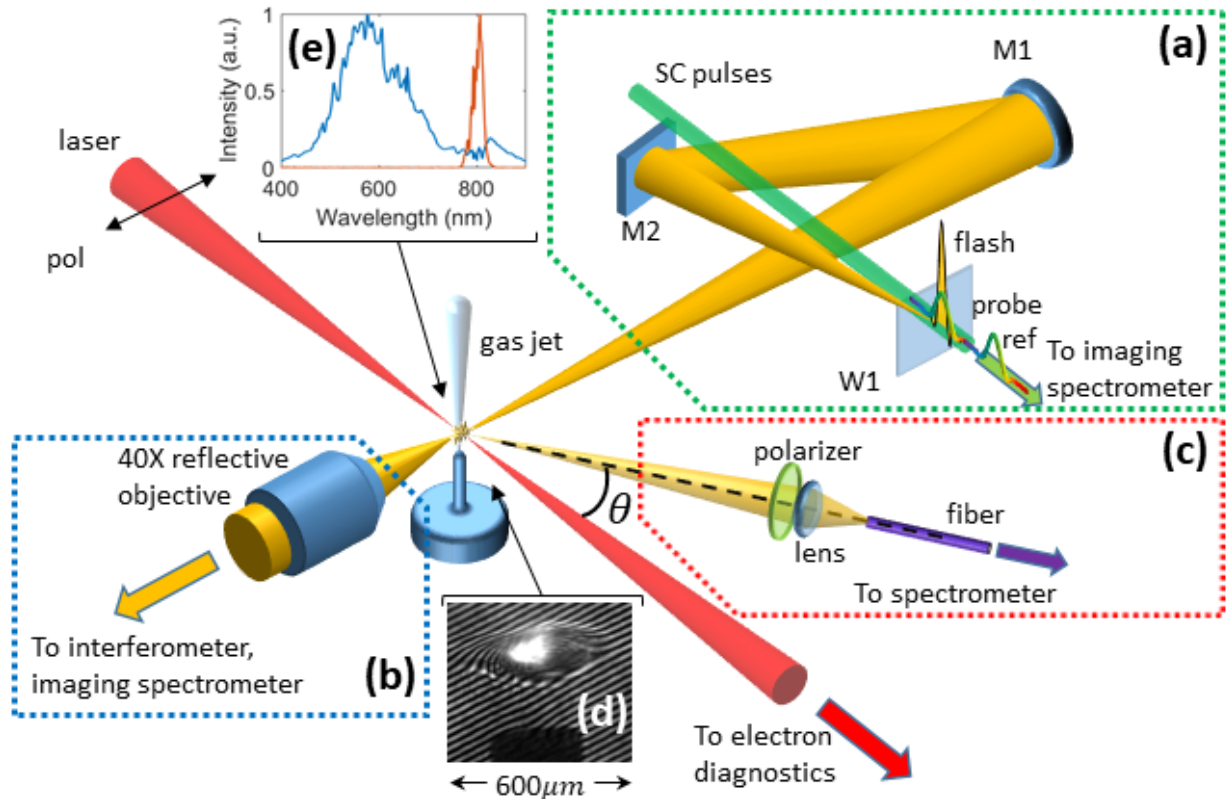

**Figure 1.** Experimental diagram showing laser-gas jet interaction and diagnostics. (a) Setup for single-shot-supercontinuum spectral interferometry (SSSI) for either direct spectral interferometry of the flash (no glass window W1) or for measurement of the flash temporal envelope using flash-induced cross phase modulation in W1. (b) 40× all-reflective objective and imaging spectrometer for high-resolution space-resolved spectra. The objective can be taken out for plasma interferometry and neutral gas density measurement using a folded wave-front interferometer (not shown). (c) Fibre-coupled spectrometer setup for flash angular distribution and polarization measurements. (d) Image of flash superimposed on interferogram of laser-plasma interaction; the flash image is brighter than the interferometer fringes. The rectangular shadow at the bottom of the image is the gas jet orifice. (e) Typical transverse (90°) flash spectrum (blue curve) and laser spectrum (brown curve) for comparison.

The experimental setup is shown in Fig. 1. Pulses of energy up to 40 mJ from a Ti:Sapphire laser ($\lambda$=800nm, 50 fs) are focused into a dense helium or hydrogen gas jet with an $f$/9.5 off-axis parabolic mirror to a full width at half maximum (FWHM) spot size $w_{FWHM}$=8 $\mu m$. In either helium or hydrogen, the plasma peak density is in the range $N_e/N_{cr} \sim 0.1 - 0.3$ with a density profile width $d_{FWHM} \sim 300 \mu m$. The neutral gas and plasma density profiles were measured using a λ=400 nm, 70 fs probe pulse (derived from the main pulse) directed transversely through the jet to a folded wavefront interferometer (not shown in the figure, but in the direction indicated in Fig. 1(b)). As the flash is produced by electron injection into relativistic plasma waves, it is always associated with the generation of forward directed relativistic electron beams. For all experiments at a given laser energy and plasma profile, flash emission was measured under conditions of maximum charge for accelerated beam energy > 125 keV, as measured by integrating the electron beam images on a LANEX scintillating screen shielded by a 100μm thick Al foil[5].

High-resolution spatio-spectral images of the flash were taken transversely using a 40× all-reflective objective and imaging spectrometer (Fig. 1(b)), while angle and polarization resolved flash spectra were recorded by fibre-coupled spectrometers (Fig. 1(c)). A typical image of the bright flash, collected along with an interferogram of the laser interaction with helium plasma, is shown in Fig. 1(d). The image and interferogram are collected by the interferometer optics (not shown) in the direction shown in Fig. 1(b). A single-shot flash spectrum collected in the same direction is shown in Fig. 1(e), with the pump laser spectrum overlaid for reference. The flash spectrum, centered at approximately $\lambda \sim 550$nm, has an extremely wide bandwidth, $\Delta\lambda/\lambda > 0.5$. The modulations in the flash spectrum are not noise—their physical significance will be discussed later in the paper.

The flash spatio-temporal evolution was measured using single shot supercontinuum spectral interferometry (SSSI), a technique developed by our group [9, 10]. The single shot capability of SSSI is essential, as the flash can vary from shot-to-shot even with nominally similar laser and gas jet parameters. We use SSSI in two ways. In method (1), radiation from a single flash (yellow beam in Fig. 1(a)) is collected at *f*/3 with 3× magnification by silver mirror M1, 0.2 *m* from the source, and directly interfered with a single well-characterized supercontinuum (SC) probe pulse at the imaging spectrometer in Fig. 1(a), with 1D space resolution 5 μm along the direction of the slit, oriented perpendicular to the laser propagation direction. In the case of multiple, spatially distributed flashes in a single shot, this resolution is sufficient to image a single flash when it appears in the field of view. The SC pulse is generated by filamentation of a small portion of the laser pulse in a 2-atm xenon cell. This measurement enables determination of the flash spectral coherence and extraction of its spectral phase. In method (2), we use a pair of SC pulses—a reference pulse and a probe pulse-- to measure the ultrashort transient Kerr effect induced by the flash in a 200 $\mu m$-thick BK7 glass window W1. Here, a single SC pulse is first generated by filamentation in the Xe cell, followed by a Michelson interferometer to generate the pair, and then followed by dispersing prisms that chirp the pulses to ~1 ps. The reference and probe pulses spatially overfill the flash image in the glass, with the reference pulse preceding the flash, and the probe temporally overlapped with it. Fourier analysis of the reference-probe interferogram at the spectrometer focal plane yields the 1D space-resolved transient Kerr phase shift in the glass, which is proportional to the flash spatiotemporal intensity profile $I_{flash}(x,t)$, where *x* is along a 1D slice in the plasma perpendicular to the laser propagation direction.

## 3. Experimental results and discussion

The spectral coherence of the flash is demonstrated in Fig. 2, which shows a broadband spatio-spectral interferogram (Fig. 2(a)) and corresponding lineout (Fig. 2(b)) from spectral domain interference between the flash and a SC probe pulse using method (1) described above. The high contrast fringes (visibility~0.62 near ~650 nm) in the $\lambda$=425-710 nm overlap region of the SC and flash spectra demonstrate spectral coherence of the flash. The curvature of the fringes arises from interference between the planar wavefronts of the SC probe and the spherical wave fronts from the point-like flash emission imaged at the source. In Fig. 2(a), the fringe period is largest near $\lambda$~535nm, which is where, in this case, the flash temporally overlaps the chirped probe pulse and then deceases at longer and shorter wavelengths. Knowledge of the spectral phase of the SC probe allows extraction of the spectral phases of single flashes, which are flat to within the uncertainty of the probe spectral phase [11].

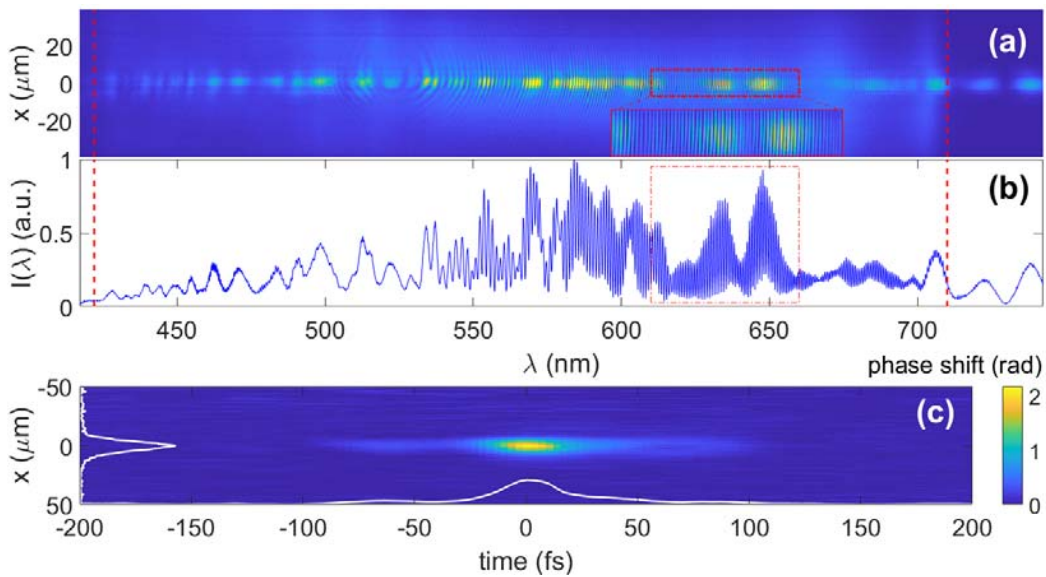

**Figure 2.** (a) Spectral interferogram of flash and supercontinuum (SC) probe pulse. Curved fringes result from interference between the point-source-like flash and the planar phase fronts of the SC probe. The centre of fringe curvature at $\lambda$~535nm is where the flash temporally overlaps the chirped probe pulse. The larger red outlined box is an expanded section of the spectrum to show the actual aliasing-free fringe spacing. (b) Lineout of (a) at z=0 $\mu m$. The vertical red dashed lines indicate the extent of the SC spectrum and the interference fringes, and the box corresponds to the smaller rectangle in (a). (c) SSSI reconstruction of the flash spatio-temporal profile. Space and time lineouts along $z$=0 and $t$=0 are shown as white curves. The spatial and temporal FWHM are ~9 $\mu m$ and ~30 fs, both limited by space and time and resolution (see text). Laser energy 30 mJ, helium plasma density $N_e/N_{cr}$ = 0.26.

Higher-resolution flash spectra, 1D space-resolved with a resolution of ~2 µm, were imaged with a 40× reflective microscope objective to the entrance slit of another imaging spectrometer (Fig. 1(b)). This setup avoids chromatic aberration of the broadband flash images. Lower resolution polarization- and angle-resolved flash spectra were collected by fibre-coupled spectrometers [11] in the range $\lambda$= 0.2 -2.5 µm (Fig. 1(c)).
The flash spatio-spectral complexity is analyzed using high-resolution images taken with the reflective microscope and imaging spectrometer setup of Fig. 1(b), with results shown in Fig. 3 for the laser vacuum focus position at the centre of the jet. Here, the imaging spectrometer has its slit aligned along the laser propagation direction (shown by white arrow in Fig. 3(a)) and records the evolution of the radiation spectrum along *z*. The pump laser energy was varied 20 - 40 mJ for peak plasma density $N_e/N_{cr} = 0.08$ or $0.16$ (1.3 or $2.7 \times 10^{20} \text{cm}^{-3}$). For our conditions, flashes are mostly emitted at two axial (*z*) locations and spectra appear either with clearly separated peaks or with more continuous features, with the latter appearing with increasing laser energy or plasma density. At lower laser energy and density (Fig. 3(a), 20 mJ, $N_e/N_{cr} = 0.08$) the spectrum at *z*~ 60μm consists of discrete peaks with equal spacing close to the plasma frequency, $\Delta\omega \sim \omega_p$=0.61 PHz at density $N_e = 1.3 \times 10^{20} cm^{-3}$, and from their frequency ratio of 4:5:6 they are identified as the 4th, 5th and 6th harmonic of $\omega_p$. This is the first observation of optical electromagnetic radiation at harmonics of the plasma frequency. Another flash occurs ~20 μm downstream in the propagation direction, where the wavelength spacing of the peaks decreases slightly and they shift to higher wavelength, consistent with locally lower plasma frequency along the density downramp. At higher density (Fig. 3(b)), the peaks are less distinct and more closely spaced, with the first flash slightly earlier in *z* and a second flash occurring ~40 μm downstream of the first.

The spectral interferograms and flash spectral phase are used to determine the time dependence of single flashes. Panels 3(a′) and 3(b′) show the square magnitude of the Fourier transforms of spectra 3(a) and 3(b) (using the extracted approximately flat spectral phase from the SSSI measurement), yielding the time dependence of the radiation intensity for the *z* position strips indicated by the black and red line segments on the left of the plots. It is seen for all cases that the flash is a sharp emission burst of duration < 5 fs followed by decaying modulations at the local plasma period $\tau_p = 2\pi/\omega_p$. At higher laser energies (panels (c′)-(f′)), the temporal modulations following the initial burst become less distinct, with more rapid decay. Also at higher energy, the spatial location of the first flash moves upstream, consistent with the earlier onset of relativistic self-focusing beam collapse. As the measured flash spectrum is limited at the high and low frequency ends by coherence and plasma absorption (see later), the initial flash burst is likely significantly shorter than 5 fs.

Among electromagnetic radiation sources from a LWFA, injection flash radiation is the only temporally coherent source measured thus far originating from the plasma accelerating structure. Another temporally coherent source is transition radiation from accelerated electron bunches exiting the plasma-vacuum boundary [12], while betatron radiation is spatially but not temporally coherent [13]. The observation of coherent emission *at* the harmonics of the plasma frequency is distinct from Raman scattering of the pump laser pulse[1,14], where spectral features would be shifted from the pump frequency by integer multiples of $\omega_p$. The physical interpretation of flash radiation at harmonics of the plasma frequency is straightforward: as the nonlinearly steepening plasma wave propagates forward at the laser group velocity, it traps and accelerates background electrons. For the case of lower laser energy and density, the plasma wake potential is still sufficiently weak to allow some of these electrons to slip back and be trapped by later successive plasma wave buckets, separated in time by $\tau_p$. As the plasma wake propagates past a fixed spatial location viewed transversely by the imaging spectrometer, the repeated trapping and acceleration of background electrons at intervals of $\tau_p$ gives rise to the modulated spectrum. From the width of the spectral peaks in Fig. 3(a), one can estimate that a train of ~9 buckets contributes to the emission, assuming the spectral interference of identical sequential optical pulses [11], which at this plasma density is >90 fs, extending well behind the 40fs laser pulse. For the larger amplitude plasma waves driven at higher laser intensity and higher density, strong trapping of background electrons occurs in the bucket closest to the laser pulse, and beam loading decreases the amplitude of successive buckets and their associated trapping.

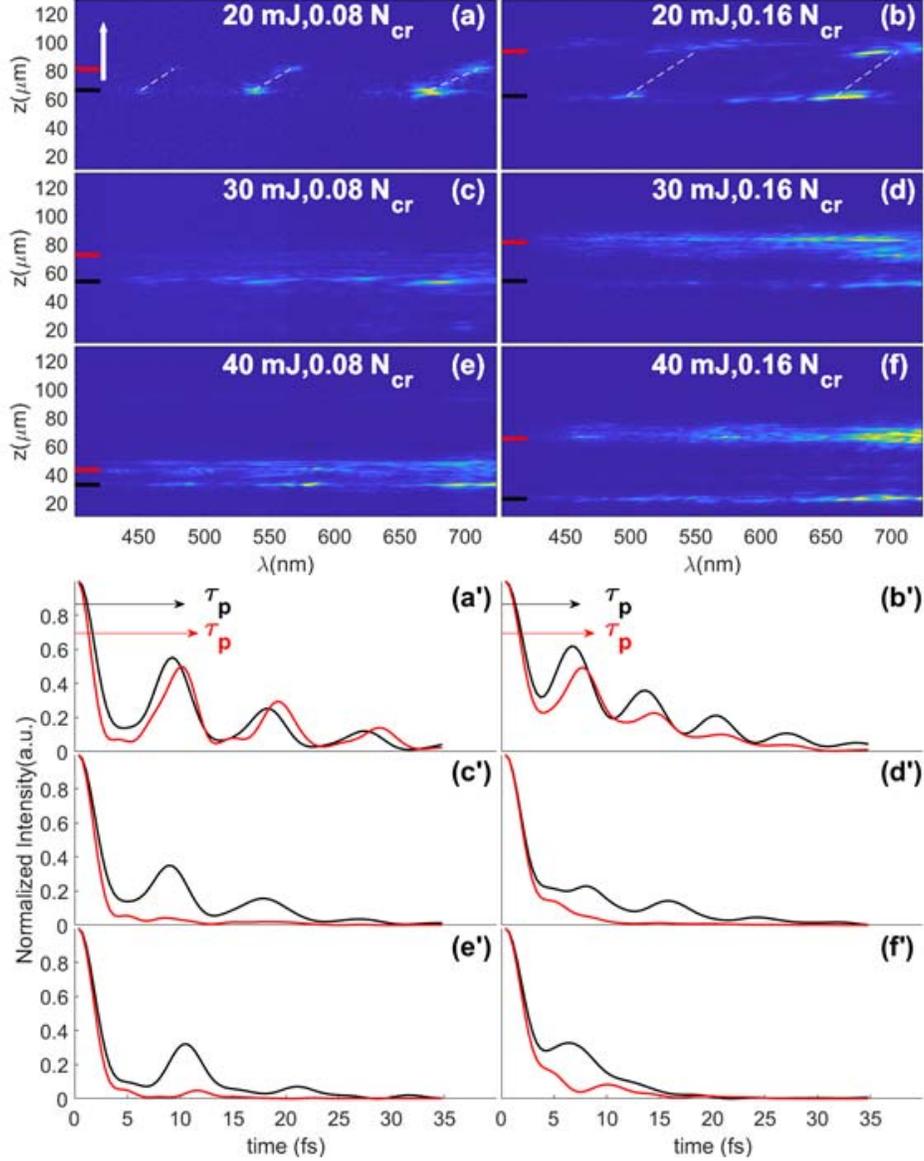

**Figure 3.** (a)-(f): Space-resolved flash spectra for laser energies 20, 30, 40 mJ and peak densities $N_e/N_{cr}$ = 0.08 and 0.16. The laser propagates in the direction of the white arrow in (a). (a′)-(f′): Square of the frequency-to-time Fourier transforms of the (a)-(f) spectra at fixed $z$ positions indicated by the back and red line segments (20 rows of CCD camera pixels), which correspond to the black and red curves. The oscillations are spaced by the local plasma period $\tau_p$. The white dashed lines in (a) and (b) show the density downramp dependence of the harmonics of plasma frequency $\omega_p$ (from left to right): (a) 6th, 5th, 4th, and (b) 4th, 3rd.

We also measure the temporal envelope of single flashes using SSSI, described above as method (2) and depicted in Fig. 1(a). The overall temporal resolution of this measurement is limited by the SC bandwidth and spherical aberration of the collection optics to $\Delta t_{min} \sim 20$ fs [11]. The spatio-temporal trace in Fig. 2(c) shows a flash emission $\tau_{FWHM} \sim 30$fs, with a long tail extending beyond ~50fs. To within the temporal resolution, this is consistent with the curves shown in Fig. 3(a′)-(f′), where the emission consists of multiple ~5 fs peaks, with fewer peaks and faster decay at higher laser energy. From Fig. 3(a) and (b), the flash extent along the propagation direction indicated by the arrow is $d_{flash}$~2.5 μm, near the imaging resolution limit, which suggests the acceleration radiation timescale is, at most , $\tau_{accel} < d_{flash}/c$ ~10 $fs$. It is remarkable that a demagnified image in the glass of only a small portion of this coherent source emission generates a nonlinear phase shift of ~2 radians (Fig. 2(c))—equivalent to flash intensity at the image of $2\times10^{12}$ W/cm$^2$ . Using this result and the measured flash angular distribution (see below), we estimate the flash energy as ~0.4 mJ with a $4\pi$ average peak brightness of ~250 J cm$^{-2}$sr$^{-1}$, conservatively assuming a resolution-limited 2μm source radius (see later discussion for Fig. 4(c)). Assuming a smaller source size from the simulation (see Sec. III) of 150nm×300nm and a ~5 fs emission burst as seen in Fig. 3 at higher density, gives an intensity brightness of ~$3\times10^{18}$ W cm$^{-2}$sr$^{-1}$. For the flash bandwidth measured, this corresponds to a brightness temperature of $T_b$~ $10^{18}$ K. As comparisons, the brightness temperature of the sun is <$10^4$ K and is in the range $10^{25}$-$10^{31}$ K for pulsars [7].

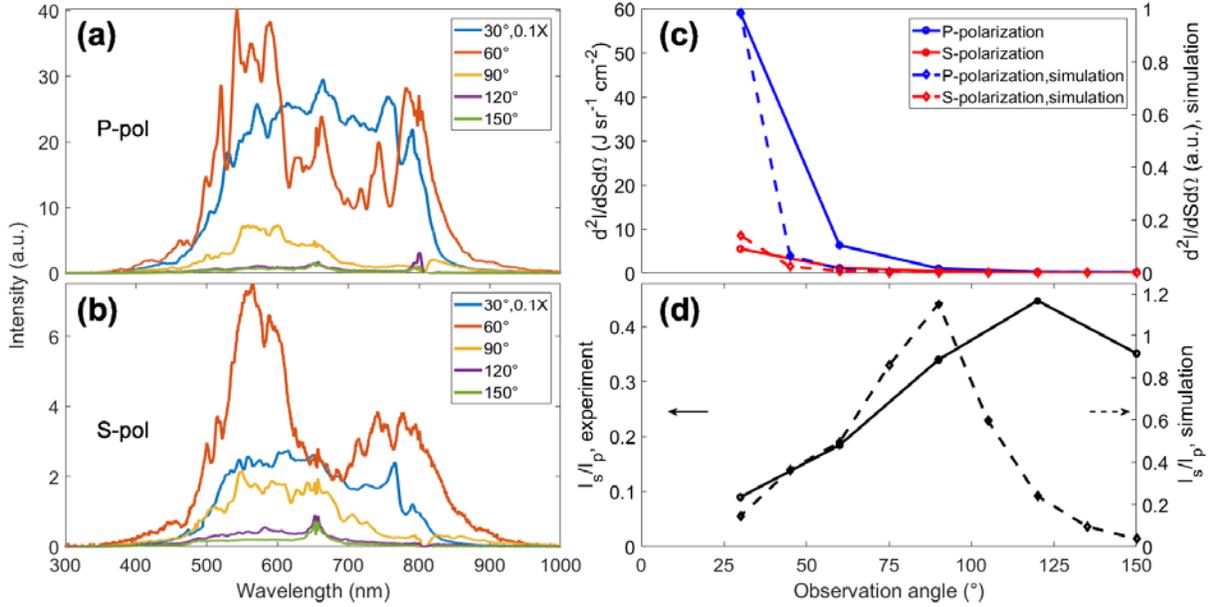

**Figure 4.** (a) (b) Spectrally resolved angular distribution of P- and S-polarized radiation. The angle in the legend is $\theta$ in Fig. 1(c). (c) Flash brightness integrated over λ=380–1000 nm for S- and P-polarization as a function of observation angle. Solid curves: experiment; dashed curves: simulation. The simulation curves are normalized to the experimental P-polarization brightness at θ=30°. (d) Ratio of S- to P-polarized radiation vs. angle $\theta$ . Solid curve: experiment; dashed curve: simulation.

The angular distribution and polarization of the flash spectrum is shown in Fig. 4 for incident laser energy 40 mJ and a hydrogen gas jet with peak plasma density $N_e/N_{cr}$~0.1 . The setup is shown in Fig. 1(c). The spectrum is collected by fibre-coupled broadband visible and IR spectrometers covering a combined range 200–2500 nm [11]. The optical fibre is mounted to a rotating arm which scans in azimuthal angle $\theta$ =30–150° with respect to the pump propagation direction, with the flash radiation passing through a broadband polarizer before entering the fibre. The collection acceptance solid angle is $\Delta\Omega = 1.8 \times 10^{-3}$ sr, set by the fibre collimator aperture, giving an angular resolution of $\delta\theta \sim \sqrt{\Delta\Omega} \sim 2.5°$. This measurement has no space resolution at the flash source; multiple flash locations contribute to the collected spectra. The flash polarization is sampled over $\phi = 0$–90°, where $\phi = 0°$ refers to P-polarization (in the plane of the pump pulse direction, the fibre arm, and the pump polarization), and $\phi$=90° describes S-polarization perpendicular to that plane. Only P-polarization (∥) and S-polarization (⊥) results are shown here, with full polarization scans shown in [11]. The fiber collection efficiency was normalized over the full angular scan range by the isotropic and unpolarized 3d-2p hydrogen recombination emission at λ=656 nm [15]. As the transmission cone angle of the laser and laser-induced SC at the jet exit is approximately $\Delta\theta_{laser} < 10°$, there was no laser contribution to the flash radiation collected at $\theta$=30°, the smallest forward angle sampled.

Examining Fig. 4(a)-(c), it is seen that for both P- and S- polarization, the flash angular distribution peaks in the forward direction and decays rapidly with increasing azimuthal angle. Under all conditions, little to no radiation is collected at $\lambda \lesssim 350\ nm$ or $\lambda \gtrsim 950\ nm$. In the forward direction, the collected radiation is dominantly P-polarized. The spectrally integrated brightness $(d^2I/dSd\Omega)_{\perp\parallel}$ of S- and P-polarized emission is plotted vs. angle in Fig. 4(c). The vertical scale is based on the collection solid angle and absolute energy calibration of the spectrometer [11], and a conservative estimate of the source area $dS$ uses the resolution-limited half-width-at-half maximum of ~2 μm from the spatio-spectral image of single flashes in Fig. 3(b). The scale is consistent with the estimate based on the nonlinear phase shift induced by the flash in glass (Fig. 2(c)). As the angle increases away from forward, there is an increasing S-polarization component, as shown in Fig. 4(d), which plots $I_s/I_p = (d^2I/dSd\Omega)_\perp[(d^2I/dSd\Omega)_\parallel]^{-1}$. As a check on the polarization sensitivity of the flash emission, we inserted a λ/2 plate to rotate the pump polarization from P to S. It is seen in [11] that the flash ratio $I_s/I_p$ (at θ=90°) follows the pump polarization.

## 4. Flash simulations and discussion

The flash spectrum, angular distribution and polarization are characteristic of synchrotron emission from laser-assisted injection of electrons into the wakefield from off-axis locations on the curved plasma wave crest. The physical picture for this emission scenario is provided by 3D particle-in-cell (PIC) simulations using the code TurboWAVE [16], with full details of the simulation described in [11]. The parameters were chosen to save computation time while allowing for multiple pulse lengths (here ~7) to fit inside the jet and for a pulse envelope longer than one plasma period (here ~2). In these respects, the laser-induced plasma wave buckets are similar to the first few following a longer pulse. Figure 5 shows simulation results for a laser pulse linearly polarized along $y$ with normalized vector potential $a_0 = 1$, pulse width $\tau_{FWHM} = 13.4$ fs and focal spot size $x_{FWHM} = 3.1 \mu m$. The plasma density profile is a 13 $\mu m$ plateau with $N_e/N_{cr} = 0.15$ and 14 $\mu m$ linear ramps on both sides. The simulation window moves at $c$ in the +z direction.

To examine the flash radiation field $B_z$ just before and after electron injection is plotted in Fig. 5(a). Before injection (inset panel), the near-planar phase fronts correspond to the relativistically self-focused laser pulse, whereas after injection there appears a short burst of radiation with spherical phase fronts. The radiation spectrum, shown in Fig. 5(d), is obtained by Fourier transforming, in the $x$ direction, the red-boxed region of panel (a). The central wavelength of ~500 nm is in qualitative agreement with the experiment. The trajectory of a trapped electron contributing to this radiation burst is shown in Fig. 5(b). An expanded view in the inset shows the initial relativistic figure eight-like motion (labeled as 1) of an off-axis plasma background electron in the plane of polarization of the laser field. The electron is first pushed to the left (2) by the $E_z$ component of the self-focused laser field and then is abruptly injected from the electron crest by the $E_y$ component (3) and trapped by the deep potential well that just precedes the crest. The trapped electron is then swept into the on-axis accelerated bunch (4). A movie tracking the electron injection is shown in [11]. The simulation reveals that off-axis injection by the laser is the dominant mechanism: For total electron displacement $\delta s(t) \approx \delta s_p + \delta s_{las}$, where $\delta s_p$ is the oscillation of a background electron in the plasma wave and $\delta s_{las}$ is the laser-induced excursion, the latter is sufficient to preferentially kick electrons from the crest, via polarization plane trajectories, into the accelerating potential well of the plasma wave. This is shown in Fig. 5(c), which plots the normalized electron velocity components $\beta_{x,y,z}$, confirming that the dominant acceleration is in the laser polarization plane (*yz*) and that the injection is laser-assisted. The strongly forward-curved plasma wave fronts [17] make this injection mechanism 3D in nature, with a less stringent dependence on electron velocity than in 1D. In 1D, there would be no laser assist and injected electrons would need to match the plasma wave's phase velocity, inducing wave breaking. By contrast, laser-assisted injection preserves the plasma wave coherence, enabling multiple successive injections as seen in the modulated spectra of Fig. 3.

The angular distribution, polarization, and frequency dependence of the flash are simulated by calculating the total field radiated from the trajectories of the $2 \times 10^4$ electrons from the PIC simulation that have ultimate post-acceleration energy of at least $\gamma - 1 > 0.001$. The polarization-, angle-, and frequency-dependent energy distribution of the injection radiation is [18]

$$\left(\frac{d^2 I}{d\omega d\Omega}\right)_{\perp,\parallel} = \frac{e^2}{4\pi^2 c}\left|\hat{\boldsymbol{\varepsilon}}^*_{\perp,\parallel} \cdot \int_{-\infty}^{\infty} dt \sum_j \frac{\hat{\boldsymbol{n}} \times [(\hat{\boldsymbol{n}} - \boldsymbol{\beta}_j) \times \dot{\boldsymbol{\beta}}_j]}{(1 - \boldsymbol{\beta}_j \cdot \hat{\boldsymbol{n}})^2} e^{i\omega\left(t - \frac{\hat{\boldsymbol{n}} \cdot \boldsymbol{r}_j(t)}{c}\right)}\right|^2, \qquad (1)$$

where $\hat{\boldsymbol{n}}$ is a unit vector pointing from the source to the observation point, $\hat{\boldsymbol{\varepsilon}}_{\perp,\parallel}$ is the unit vector for S- and P-polarization, and $\boldsymbol{r}_j, \boldsymbol{\beta}_j$ and $\dot{\boldsymbol{\beta}}_j$ are the position, normalized velocity, and acceleration of electron *j*, and where the sum is over the $2 \times 10^4$ electrons. The simulation results explain the measured flash angular distribution and polarization dependence, as well as the coherence. The spectrally integrated source brightness $(d^2 I/dSd\Omega)_{\perp,\parallel}$ of S- and P-polarized radiation is plotted in arbitrary units as a function of $\theta$ in Fig. 4(c) and agrees with the measurements: As the off-axis electron injection to relativistic velocities is dominantly forward directed, the radiation associated with this acceleration will also be forward-directed, accounting for the forward-peaked angular distributions. The dominance of P-polarization of the flash is a direct signature of the laser-polarization-plane orbits favoured by laser-assisted injection. By contrast, for the case of wave-breaking-dominated injection, there would be no preferred flash polarization as there would be no preferred electron trajectories.

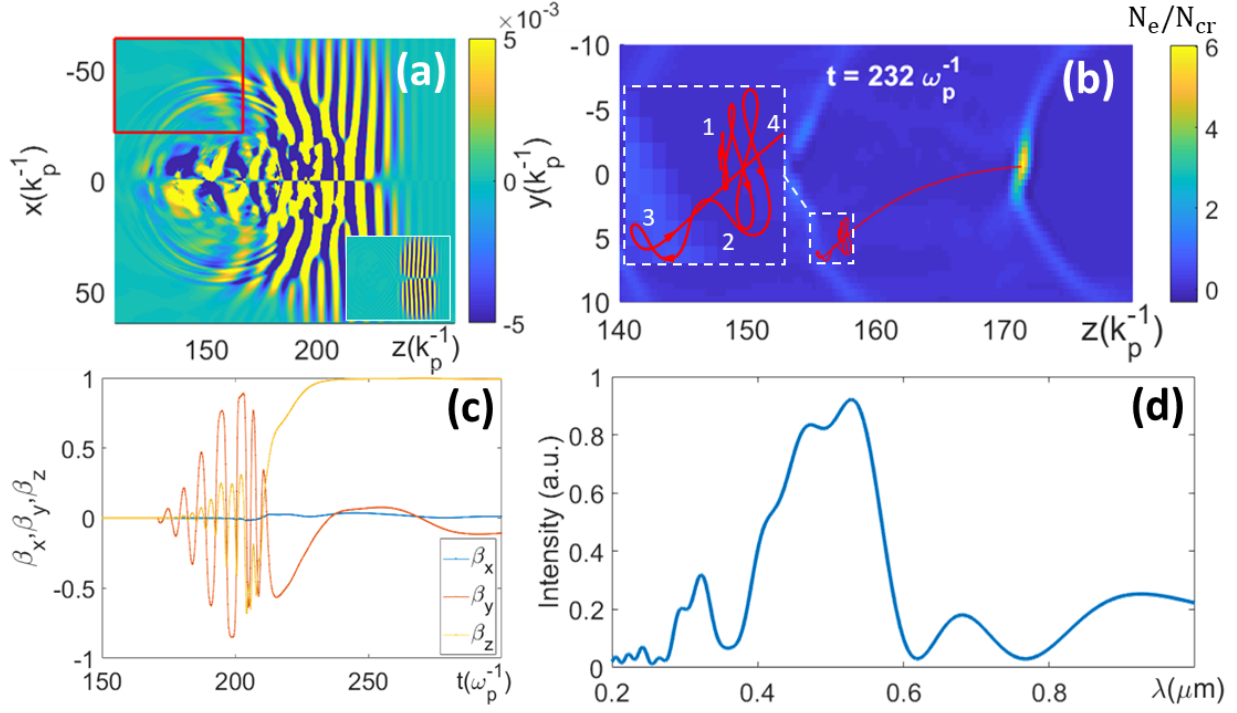

**Figure 5.** Results from 3D particle-in-cell simulation. The laser pulse propagates from left to right. (a) $B_z$ before (inset) and after (main panel) electron injection. The red box encloses a portion of the spherical wave radiated by the injected electron bunch. (b) Lab frame trajectory of electron trapped and accelerated by plasma wave bucket. Orbit phases (1) – (4) are described in the text. Forward curved light blue features are electron crests of the relativistic plasma wave. (c) Normalized velocity components of the electron tracked in panel (b), showing laser-assisted injection. (d) Square of Fourier transform (in *x*-direction) of spherical wave radiation in red box of (a).

As in the experiment, the simulation of $I_S/I_P$ (Fig. 4(d)) shows an increasing contribution of S-polarized radiation at transverse observation angles, except that at its peak, $I_S/I_P$ is <1 in the experiment and >1 in the simulation. The peaking of $I_S/I_P$ at transverse observation angles is likely an effect of coherence, as discussed below.

The observation angle-dependent flash spectrum is also computed using Eq. (1) and is shown in [11]. It is seen that emission harmonics at the local plasma frequency are evident for transverse observation angles but less clear for forward or backward angles. This is consistent with our transverse observation (Fig. 3) of a source modulated at $\tau_p$ at a fixed spatial location. Observation at non-transverse angles is susceptible to either or both of inter-flash (in the case of several flash spatial locations) or intra-flash spectral interference.

Aside from its detailed spectral dependence on the injected electron trajectories as discussed above, the flash's approximate overall wavelength range can be inferred from considerations of coherence and absorption. The effect of coherence is described by the structure factor $F(\mathbf{q}) = \left|\int d^3\mathbf{r}\, N_{e,rad}(\mathbf{r}) e^{i\mathbf{q}\cdot\mathbf{r}}\right|^2 \propto N_{e0}^2 e^{-\frac{1}{2}q^2(w^2 \cos^2\theta + a^2 \sin^2\theta))}$ applied to injection radiation originating from trapping at the plasma wave crest. Here $\mathbf{q}$ is the radiation wavevector and we take the distribution of radiating electrons through and across the crest as $N_{e,rad}(r) = N_{e0} e^{-z^2/w^2 - \rho^2/a^2}$, where the crest FWHM thickness from the simulation is $w_{fwhm} = 2\sqrt{\ln 2}\, w \sim 150 nm$ and the bunch width is $a_{fwhm} = 2\sqrt{\ln 2}\, a \sim k_p^{-1} \sim 300 nm$ (see Fig. 5(b)). The structure factor predicts a short wavelength $1/e^2$ "coherence cutoff" at $\lambda_{short} \sim 300{\sim}400 nm$ over a wide range of $\theta$, in reasonable agreement with the experimental spectra . The long wavelength cutoff is set by the electron density in the crest, which at the time of injection peaks at $N_e/N_{cr} \sim 1 - 1.3$ in the simulation, giving a long wave cutoff $\lambda_{long} = 850 - 950 nm$, also in reasonable agreement with the measured spectra.

While the laser-assisted injection mechanism ensures that S-polarized flash emission is weak at all angles θ, its relative contribution $I_S/I_P$ peaks at transverse angles as seen in the experimental and simulation plots of Fig. 4(d). This can be explained as a consequence of coherence by considering the classes of electrons likely to radiate either more S-polarized radiation or more P-polarized radiation. The "class P" electron orbits originate in the laser polarization plane (*yz*), while class S orbits contributing to S-polarized radiation must originate close to the *xz* plane. At observation angle θ ~ 90°, $F_{P,S}(\mathbf{q}) \propto N_{e0,P,S}^2 e^{-\frac{1}{2}q^2 a_{P,S}^2}$, where $F_{P,S}$ is the structure factor applied to class P or class S electrons, $N_{e0,P,S}$ is the effective local electron density of injected class P or S electrons, and $a_{P,S}$ is the extent of their transverse distribution as viewed from θ~90°. Therefore, we take $a_S \sim 0$ and $a_P \sim k_p^{-1}$, and a rough estimate gives $I_S/I_p \sim F_s/F_p \sim (N_{e0,S}/N_{e0,P})^2 e^{\frac{1}{2}(\lambda_p/\lambda)^2}$, where $\lambda_p$ is the plasma wavelength and $\lambda$ can be taken as the peak flash wavelength. Using parameters from the simulation, $N_{e0,S}/N_{e0,P} \sim 0.1$ and $\lambda_p/\lambda \sim 3$, gives $I_S/I_p \sim 1$ , in qualitative agreement with the experiment and simulation results.

## 5. Conclusions

In conclusion, we have measured in detail and characterized for the first time spectrally coherent radiation generated directly from a laser wakefield accelerator. This coherent, ultra-broadband, bright, and ultrashort pulse radiation flash originates from laser-assisted injection of electrons into the accelerating structure of the relativistic plasma wave. The flash is sufficiently bright to induce large nonlinear refractive index shifts in optical materials. Analysis of this injection radiation provides a detailed picture of the dynamics inside the laser plasma accelerator. We conclude that the injection process does not occur from pure wave-breaking: electrons are kicked into the accelerating phases of the plasma wave by the laser field. Under some conditions, periodic injections at the plasma period, which give rise to the first observation of coherent radiation at harmonics of the plasma frequency, show that the plasma wave can remain coherent through the injection process. The initial burst of injection flash radiation was found to be no longer than approximately a single optical cycle in the visible spectrum.  However, our measurements were limited by our probe pulse bandwidth and the low and high frequency cutoffs of the flash spectrum. It is likely that the flash burst duration is considerably shorter.

**Acknowledgements.** This research was supported by the National Science Foundation (PHY1619582), U.S. Department of Energy (DESC0015516), U.S. Department of Homeland Security (2016DN077ARI104).